# Perceptual similarity of visual patterns predicts dynamic neural activation patterns measured with MEG

Susan G. Wardle[1], Nikolaus Kriegeskorte[2], Tijl Grootswagers[1], Seyed-Mahdi Khaligh-Razavi[2], and Thomas A. Carlson[1,3,]*

[1]Department of Cognitive Science and ARC Centre of Excellence in Cognition and its Disorders and Perception in Action Research Centre, Macquarie University, Sydney, New South Wales, 2109, Australia

[2]Medical Research Council, Cognition and Brain Sciences Unit, Cambridge, CB2 7EF, UK

[3]Department of Psychology, University of Maryland, College Park, MD, USA

*Corresponding Author:
Thomas A. Carlson
Department of Cognitive Science
Australian Hearing Hub
16 University Avenue
Macquarie University NSW 2109 Australia
Phone : +61 2 9850 4133
Email: thomas.carlson@mq.edu.au

Acknowledgements. This research was supported by an Australian NHMRC Early Career Fellowship (APP1072245) awarded to S.G.W., a European Research Council Starting Grant (ERC-2010-StG 261352) awarded to N.K., and an Australian Research Council Future Fellowship (FT120100816) awarded to T.A.C. The authors declare no competing financial interests.

**Abstract**

Perceptual similarity is a cognitive judgment that represents the end-stage of a complex cascade of hierarchical processing throughout visual cortex. Previous studies have shown a correspondence between the similarity of coarse-scale fMRI activation patterns and the perceived similarity of visual stimuli, suggesting that visual objects that appear similar also share similar underlying patterns of neural activation. Here we explore the temporal relationship between the human brain's time-varying representation of visual patterns and behavioral judgments of perceptual similarity. The visual stimuli were abstract patterns constructed from identical perceptual units (oriented Gabor patches) so that each pattern had a unique global form or perceptual 'Gestalt'. The visual stimuli were decodable from evoked neural activation patterns measured with magnetoencephalography (MEG), however, stimuli differed in the similarity of their neural representation as estimated by differences in decodability. Early after stimulus onset (from 50ms), a model based on retinotopic organization predicted the representational similarity of the visual stimuli. Following the peak correlation between the retinotopic model and neural data at 80ms, the neural representations quickly evolved so that retinotopy no longer provided a sufficient account of the brain's time-varying representation of the stimuli. Overall the strongest predictor of the brain's representation was a model based on human judgments of perceptual similarity, which reached the limits of the maximum correlation with the neural data defined by the 'noise ceiling'. Our results show that large-scale brain activation patterns contain a neural signature for the perceptual Gestalt of composite visual features, and demonstrate a strong correspondence between perception and complex patterns of brain activity.

Keywords: magnetoencephalography (MEG), representational similarity analysis, perceptual similarity, representational geometry, decoding, Gestalt perception

**Introduction**

Judgments of perceptual similarity require integrating information across a complex hierarchical network of brain regions. An early idea of how perceptual similarity might be conceived at the neural level is as a product of representational distance (Shepard, 1964; Torgerson, 1965). Specifically, visual objects that appear similar are assumed to share similar underlying neural representations. One of the first demonstrations of this idea with fMRI showed that different object categories (such as faces, houses, chairs) that share image-based attributes also share a similar underlying neural structure (O'Toole et al., 2005). Similarity in stimulus structure and in brain activation patterns for object categories were both defined by a classification analysis on the principal components derived from either the stimulus set or the patterns of fMRI activation; and categories that were more confusable with image-based classification were also more confusable in their brain activation patterns.

Building on this mapping between stimulus similarity and neural representation, several studies have observed a correlation between behavioral similarity judgments for objects and their corresponding neural representations. Rotshtein et al. (2005) used morphs between famous faces within an fMRI adaptation paradigm and found that different brain regions associated with face processing were responsive to the physical features of faces (inferior occipital gyrus) versus the perceived identity of faces (right fusiform gyrus). Several studies have used rich image sets (such as objects from multiple categories) and shown that stimuli that are rated more similar by human observers also share more similar patterns of fMRI activation (Edelman et al 1998; Hiramatsu et al 2011, Mur et al 2013; Connolly et al. 2012). These results suggest that objects that appear more similar have more similar brain representations; however, since these studies have focused on object recognition, they have used stimuli in which perceptual similarity is unavoidably conflated with conceptual similarity. Other studies have emphasized the role of image statistics, and used naturalistic stimuli varying on both semantic and visual dimensions (Hiramatsu et al., 2011), in which the mapping between different feature dimensions and perceptual similarity is complex. Consequently, in these experiments it is not possible to separate out the effects of perceptual similarity from other forms of similarity among the stimulus classes.

A notable exception is a series of studies examining fMRI activation patterns for novel shapes and objects in the object-selective lateral occipital complex (LOC). In an early demonstration, Kourtzi and Kanwisher (2001) found that following adaptation, the BOLD response in LOC for stimuli with the same shape was reduced compared to that for different shapes, even when the local contours of the 'same shape' condition were changed through manipulations in stereoscopic depth and occlusion. This suggests that stimuli with similar perceived shape have more similar activation patterns in LOC, irrespective of differences in local contours. Similarly, Haushofer et al. (2008) reported that fMRI activation patterns in the anterior LOC (pFs) for novel two-dimensional shapes that varied parametrically in aspect ratio and skew correlated with the results of a same-different task with human observers; shapes that were more confusable have more similar activation patterns. Conversely, activation patterns in the posterior LOC (LO) correlated more with the physical parameters of the stimuli (i.e., the absolute magnitude of difference in aspect ratio and skew, rather than perceived shape similarity). Op de Beeck, Torfs and Wagemans (2008) reported a significant correlation between the similarity of fMRI activation patterns in LOC and ratings of perceived shape similarity for novel categories of objects defined by their shaded three-dimensional shape. In contrast to Haushofer et al. (2008), Op de Beeck et al. (2008) observed the correlation with perceptual similarity across LOC, which the authors attribute to differences between the studies in both the stimuli and the similarity task.

In sum, there is substantial evidence that the similarity of coarse-scale fMRI activation patterns can be related to the perceived similarity of visual objects of varying complexity (e.g. Op de Beeck et al., 2008; Haushofer et al., 2008; Edelman et al 1998; Hiramatsu et al 2011, Mur et al 2013; Connolly et al. 2012). The aims of the present study are to build on this work by examining the extent to which perceptual similarity is accessible in dynamic large-scale brain activation patterns measured with MEG, and to probe the structure of the underlying neural representation by comparing the temporal performance of several models. In order to separate perceptual similarity from other forms (e.g. conceptual or semantic), we use a set of abstract visual patterns as stimuli (see description below) and compare the performance of models of early visual processing and stimulus properties to a model of perceptual similarity. Most studies examining representational geometry have used fMRI (e.g. Clarke and Tyler, 2014; Edelman et al., 1998; Hiramatsu et al. 2011; Mur et al., 2013), and focused on the transformation of the representational space across spatial networks of brain regions. Compared to other neuroimaging methods, fMRI has limited temporal resolution, and consequently the temporal evolution of the mapping between behaviorally relevant features and the structure of neural representations has remained largely unexplored. To complement previous fMRI results, our focus here is on the temporal (rather than spatial) evolution of the neural representational geometry in response to visual patterns.

In order to investigate the information content of the brain's time-varying representation of the stimuli, we employed representational similarity analysis (RSA; Kriegeskorte and Kievit, 2013) to test several candidate models of the representational structure, including a model of perceptual similarity. RSA is a model-testing approach for studying brain activation patterns, which builds on traditional brain 'decoding' methods (e.g. multivariate pattern analysis) to facilitate conclusions about the content of decodable information (Kriegeskorte and Kievit, 2013). The intuition behind RSA is that differences in the decodability of stimuli can be interpreted as a proxy for neural representational similarity. Stimuli that are more difficult to decode from each other are assumed to have more similar underlying neural representations. If a model successfully predicts the representational distance between stimuli, it provides evidence that the source of representational information in the model is present in the neural population code. An additional strength of applying RSA to MEG data is that the fine-scale temporal resolution of the neuromagnetic signal reveals the emergence of representational geometry over time, providing a more complete characterization of the model's performance.

In order to systematically decouple perceived similarity from both semantics and lower-level visual features, we used an abstract stimulus set of visual patterns constructed from arrangements of Gabor patches. These stimuli will drive the response of neurons in early visual cortex, and make straightforward predictions for a range of models that can be used to characterize the evoked cortical response to the stimuli. The stimulus set varied along three dimensions: the number of elements, the local orientation of each Gabor patch, and the degree of orientation coherence among the elements. Critically, although the stimuli are constructed from identical elements, each stimulus has a unique global form or perceptual 'Gestalt' (Figure 1A). The advantage of this stimulus set is that models of early visual processing and stimulus features can easily be constructed for comparison with a higher-level perceptual RDM based on the unique global form produced by the different arrangements of Gabors. We compare a perceptual similarity model derived from ratings of the stimuli made by human observers to several models[1] based on the neural processing of low-level visual features: (1) a model based on differences in retinotopic stimulation between the stimuli, (2) a V1-like model based on HMAX (Riesenhuber and Poggio, 1999; Serre and Riesenhuber 2004; Hubel and Wiesel, 1965), (3) a model of local orientation differences between the stimuli, and (4) a model which predicts decodability based on inter-stimulus differences in the radial bias (e.g. Schall et al., 1986; Sasaki et al., 2006).

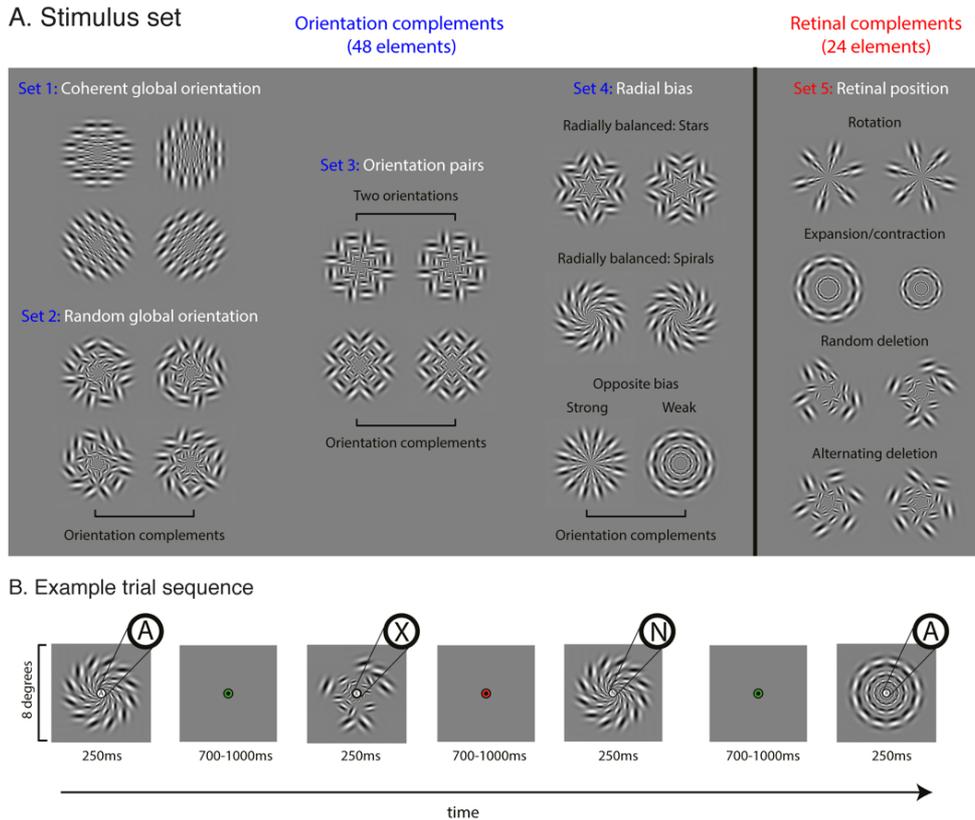

**Figure 1.** Experimental Design. (A) Visual stimuli in set 1 have a coherent global orientation [0°, 90°, 45°, or 135°], while the patterns in set 2 have an equivalent overall local orientation disparity but lack a coherent global orientation. Set 2 patterns were created by generating an array of elements with random orientations, and then rotating the elements of the random seed pattern by 90°, 45°, and 135°. In set 3, each pattern has alternating elements of two orientations (top pair: 0 and 90°, bottom pair: 45° and 135°), with the order of orientations swapped between the members of each pair (top and bottom rows). In set 4, the star and spiral pairs are radially balanced, with elements rotated either 45° or -45° relative to (invisible) radial spokes originating from fixation. The third pair contains one pattern with a strong

---

[1] We use the broad definition of 'model' implied by the Representational Similarity Analysis framework, as any potential explanation for the variance in the similarity of the brain representations observed for the visual stimuli — hypotheses which may be based on e.g. computational models, behavioral ratings, or straightforward predictions based on shared stimulus features.

radial bias (radial spokes) and one with a weak bias (rings). All pairs in set 5 are retinal complements, with the Gabor patches in complementary retinal locations.

**Materials and Methods**

*Participants*. Twenty volunteers (8 male, 12 female) with an average age of 21.6 years participated in the experiment and received financial reimbursement. Informed written consent was obtained from each volunteer prior to the experiment, and all experimental procedures received approval from the institutional ethics committee at the University of Maryland.

*Stimuli*. Visual stimuli were arrays of Gabor patches (sine wave convolved with a 2D Gaussian window) in a log polar arrangement (inner radius: 1°, outer radius: 8°) with four rings and twelve spokes (Figure 1A). The size of the elements was log scaled based on their position relative to central fixation to account for cortical magnification in early visual cortex. The 26 visual stimuli were designed in 13 complementary pairs to facilitate pairwise multivariate pattern classification as a foundation for RSA. Nine stimulus pairs were *orientation* complements constructed from 48 individual Gabors (Figure 1A, sets 1-4). In each pair, elements at corresponding spatial locations were rotated 90°. These patterns were thus maximally different in terms of orientation disparity, but equivalent in terms of coarse scale retinal stimulation. The remaining four pairs were *retinal* complements, constructed from 24 individual Gabors (Figure 1A, set 5). For these pairs, elements present in one pattern were absent in the corresponding spatial location of its complement. Four additional visual stimuli were also presented during the experiment. Due to a coding error in stimulus generation, these stimuli were either identical/ redundant in the experimental design (i.e. duplicate spiral and ring patterns). The data was not analyzed for these patterns.

*Procedure*. Participants viewed the visual stimuli while lying supine in a magnetically shielded recording chamber. Stimuli were projected onto a translucent screen located approximately 30cm above the participant. The experiment was run on a Dell PC desktop computer using MATLAB (Natick, MA, U.S.A.) and functions from the Psychtoolbox (Brainard, 1997; Pelli, 1997; Kleiner, et al. 2007). The visual stimuli were displayed on the screen in the MEG for 250ms with a variable inter-stimulus interval (700 -1000ms). Participants ran eight blocks of trials of approximately seven minutes in length, which each contained six presentations of each visual stimulus, presented in random order (48 presentations total per stimulus). Participants performed a fixation task during the experimental runs (Figure 1B), which involved detecting whether a small letter (0.5°) in the center of the stimulus was a vowel or a consonant (randomly drawn from the set {'A' 'E' 'I' 'O' 'U' 'R' 'N' 'X' 'S' 'G'}). Feedback was provided by changing the color of the fixation target for 300ms after each trial, and a performance summary was displayed after each block of trials. The mean accuracy across participants for the task was 97% correct ($SD = 2.6\%$).

*MEG Acquisition and Preprocessing*. Neuromagnetic recordings were acquired with a whole-head axial gradiometer MEG system (KIT, Kanazawa, Japan). The system had 157 recording channels with 3 reference channels. Recordings were filtered online from 0.1 to 200 Hz using first order RC filters and digitized at 1000 Hz. Time shifted principal component analysis (TSPCA) was used to denoise the data offline (de Cheveigne and Simon, 2007). Trials were epoched from -100ms to 600ms relative to stimulus onset. Trials with eye movement artifacts were removed automatically using an algorithm that detects large deviations in the root mean square (RMS) amplitude over 30 selected eye-blink sensitive channels. The average rejection rate was 2.2% ($SD = 1.0\%$) of trials across participants. After artifact rejection, the data were resampled to 200Hz, and corrected for the latency offset introduced by resampling. Principal component analysis was used to reduce the dimensionality of the data. Using a criterion of retaining 99% of the variance, the number of dimensions was reduced from 157 (recording channels) to 62 principal components, on average across subjects.

*Pattern Classification*. We used a naïve Bayes implementation of linear discriminant analysis (LDA; Duda et al., 2001) for the decoding analysis. The input to the classifier was the factor loadings for the principal components. Generalization of the classifier was evaluated using *k*-fold cross validation with a ratio of 9:1 training to test. To improve the signal to noise ratio, trials were averaged in pseudo trials (Isik et al., 2014; Meyers, 2013). Each pseudo trial was an average of four trials. The set of 48 trials per pattern (sometimes less after artifact rejection) was reduced to 10 pseudo trials by averaging a random selection of trials. Nine of the pseudo trials were used to train the classifier, and one was used to test the classifier. Thus for each pairwise comparison there were 18 trials used to train (nine from each stimulus pattern) and two used to test the classifier (one from each pattern). This procedure was repeated 100 times, each time with a new randomization. Classification accuracy is reported as average classifier performance (*d*-prime). The decoding analysis was run for all possible pairwise comparisons between stimulus patterns for each time point.

*Model Definitions*.

Within the RSA framework, we constructed several model representational dissimilarity matrices (RDMs) based on stimulus properties that may account for the decodability of the stimuli to compare with the empirical time-varying MEG RDM. Each model makes predictions about the decodability of the visual stimuli for each pairwise stimulus comparison (exceptions noted below in model definitions). The models are not intended to be comprehensive models of neural processing, but instead are used to identify what stimulus properties may underlie their decodability from the neuromagnetic signal measured with MEG. In each case the model predictions are represented as RDMs with values normalized to range from 0 (identical in terms of the model) to 1 (extremely different in terms of the model).

**Perceptual Model**. Fifty participants provided ratings of the perceived similarity of the patterns in an online study conducted using Amazon's Mechanical Turk services. Participants were briefly shown (duration: ~250ms) two of the individual patterns simultaneously and rated the similarity of the patterns on a scale from 1 to 100. The written instructions to participants read: "Judge the visual similarity of the images: Your task will be to rate how similar two abstract images are on a scale from 0 to 100 using the slider. Don't think about the task too hard, we are interested in your immediate first impression. The images will only be shown briefly and you will only get one chance to see them, so make sure that you are ready when you press the "Begin experiment" / "View next" button." Each participant made ratings for all the possible pairwise comparisons (325 comparisons total), and these were used to create a perceptual representational dissimilarity matrix (RDM) for each participant. As the visual patterns were all constructed from identical Gabors, we assume that participants based their similarity judgments for each pair on the overall global arrangement of the Gabors in each pattern. Individual RDMs were normalized to range from 0 to 1, and then averaged to create a group perceptual model RDM (Figure 4B).

**Retinal Envelope Model.** Previously, we have shown that differences in retinal projection between higher-level object stimuli are a robust predictor of their decodability with MEG (Carlson et al., 2011). To evaluate the role of retinal projection in the representational geometry of the current lower-level visual stimuli we constructed a model that predicts decodability based solely on differences between exemplars in terms of coarse scale retinotopic stimulation (Figure 4B). Specifically, this model predicts that pairs of stimuli which have individual Gabors in *different* spatial locations (retinal positions relative to central fixation) relative to each other (e.g. Figure 1, pairs in Set 5) will be easier to decode than pairs that have individual Gabor elements in spatially corresponding locations (e.g. Figure 1, Sets 1-4). Thus the retinal envelope model predicts decodability solely on the basis of differences in local retinal position between stimulus pairs.

To compute dissimilarity of their retinal position, each element location in the stimulus is a location in a vector; and at each location in the vector a 1 or 0 indicates the presence or absence of a Gabor patch. The dissimilarity between two stimulus patterns is the absolute difference between the two patterns' vectors. Dissimilarity was computed for all possible pairwise comparisons between the patterns to create the model RDM. In detail, according to this model, stimulus pairs in which both patterns have 48 elements are predicted to be difficult to decode from each other because they both have the same number of elements in the same locations (blue region in the retinal envelope model RDM shown in Figure 4B). In contrast, stimulus pairs in which one pattern has 24 elements and the other has 48 elements are predicted to be easier to decode (grey region in the model RDM). Finally, stimulus pairs in which both patterns have 24 elements but in different spatial locations (i.e.: no overlap in the position of the elements between the two members of the pair) are predicted to be the easiest to decode (red/yellow region in the model RDM). Another way of conceptualizing the stimulus differences captured by the retinal envelope model is in terms of local contrast. Pairs of patterns that have Gabor elements in different locations also have a corresponding difference in local contrast (e.g. between the mid-grey of the background in one pattern and the white-black of the Gabor in the other pattern), which is likely to contribute to decodability. RMS contrast is known to influence the overall magnitude of activation at a population level in both BOLD fMRI (Olman et al., 2004; Rieger et al., 2013) and MEG (Rieger et al., 2013)."

**V1 Model (HMAX-S1).** To approximate the response of early visual areas to the stimuli, we employed the HMAX model. The S1 layer of HMAX encodes orientation at multiple scales, based on knowledge of receptive field properties of neurons in early visual areas (Riesenhuber and Poggio, 1999; Serre and Riesenhuber 2004; Hubel and Wiesel, 1965). The dissimilarity between the visual stimuli for HMAX's S1 layers was computed using code available on the web (http://cbcl.mit.edu/jmutch/cns/index.html#hmax). The inputs to HMAX were the images of the visual stimuli (rendered at 600 x 600 pixel resolution). HMAX returns a feature vector, which represents the simulated cortical response to the stimulus. To compute dissimilarity between the stimuli, we computed the Euclidean distance between the feature vectors for each stimulus pair. Dissimilarity was computed for all possible pairwise comparisons between the stimuli to create the V1 model RDM (Figure 4B).

**Orientation Disparity model.** The orientation disparity model predicts decodability based on local orientation differences between the stimuli (Figure 6A). Orientation disparity was computed by summing the absolute orientation difference between corresponding Gabor elements in each stimulus pair. Dissimilarity was computed for all possible pairwise comparisons between the stimuli and then normalized to create the model RDM. Note that this model only makes predictions for the decodability of patterns with all of the 48 elements (Figure 1, Sets 1-4), as it is not possible to compute orientation disparity for unpaired Gabor patches.

**Radial Preference model.** Neurophysiological studies have observed a bias in the number of neurons representing radial orientations (i.e. orientations that point toward the fovea; Levick and Thibos, 1982; Leventhal and Schall, 1983; Schall et al., 1986), and this bias has also been observed in human fMRI studies (Sasaki et al., 2006; Mannion et al., 2010; Alink et al., 2013). The radial preference model predicts decodability based on inter-stimulus differences in the radial bias (Figure 6A). We modeled the radial bias in the stimuli by computing each element's orientation disparity relative to the radial orientation for its location in the visual pattern relative to fixation ($\theta$), and taking its cosine (e.g. 0° disparity = 1, 90° disparity = 0). The difference in the radial bias between two patterns was calculated as the sum of the absolute value of the difference between the radial bias responses for their spatially corresponding Gabor elements. Note that this model also only makes predictions for the decodability of patterns with all of the 48 elements (Figure 1, Sets 1-4).

*RSA Model Evaluation and Noise Ceiling*. We used the RSA framework (Kriegeskorte et al., 2008; Nili, et al., 2014) to study the brain's emerging representation of the stimuli by comparing the models to time resolved MEG RDMs (Cichy et al., 2014; Redcay and Carlson, 2015). Correspondence between the empirical RDM (MEG data) and the normalized model RDMs was assessed by computing Kendall's tau-a (i.e., a rank-order correlation) for each time point and each subject, producing a time-varying correlation between the model and MEG data. Significance was assessed with a non-parametric Wilcoxon signed rank test (FDR < 0.01). A cluster threshold of 3 consecutive time points was used to determine onset latencies at the group-level (Figure 4D). Individual subject latencies (Figure 4C) were computed by comparing each timepoint's correlation (Tau-a) to a null distribution of correlations, which were derived from bootstrapping by shuffling the RDMs (significance assessed at FDR < 0.05). The onset was computed as the first significant time point > 0 (stimulus onset). No consecutive time point criterion was used for individual latencies. Peak latencies were computed as the highest value of the correlation between the data and the model.

We used the 'noise ceiling' as a benchmark for model performance (Nili, et al., 2014). The noise ceiling estimates the magnitude of the expected correlation between the true model RDM and the MEG RDM given the noise in the data. The upper bound is calculated by correlating the group average MEG RDM with the individual RDMs. This correlation is overfitted to the individual RDMs and produces an upper estimate of the true model's average correlation. The lower bound is calculated by the 'leave-one-subject-out' approach, so that each subject's individual RDM is correlated with the RDM for all remaining subjects, preventing overfitting. The average correlation across all iterations of this calculation underestimates the correlation with the true model and defines the lower bound of the expected correlation with the true model RDM.

*Projection of weight maps onto sensor space.*

To identify the contribution of different sensors to decoding performance we constructed weight maps for four key time points: 40 ms (decoding onset), 60 ms, 90 ms (peak decoding), and 145 ms (peak correlation between perceptual RDM and MEG RDM). For each subject (*N*=20) and pairwise stimulus comparison made by the classifier *(N*=325) we bootstrapped (10x) the process of averaging 4 randomly selected trials per exemplar into pseudotrials (as used for classification, see above), and extracted the LDA weights (i.e. linear coefficients) for the comparison. As raw classifier weights are difficult to interpret (see Haufe et al., 2014), before projection into sensor space we transformed the weights using the recently described method of Haufe et al. (2014). We multiplied the weights (*W*) by the covariance in the pseudotrials so that $W' = \Sigma(\text{pseudotrials}) \times W$. The transformed weights (*W'*) for all pairwise comparisons were averaged per subject and multiplied by the subject-specific PCA coefficients to obtain a projection onto the sensor space. Next, FieldTrip (Oostenveld et al., 2011) was used to transform the gradiometer topography into the planar gradients, which were then combined and interpolated at the sensor locations to create intuitive topographic maps (Figure 3; Movie 1).

**Results**

*Early Decoding of Abstract Visual Patterns from MEG*

Recent MEG decoding studies have shown that early visual feature representations (e.g., retinotopic location, orientation, and spatial frequency) and higher-level object categories can be decoded from neuromagnetic recordings (Carlson, et al. 2011; Carlson et al., 2013; Cichy et al., 2014; Cichy et al., 2015; Ramkumar, et al. 2013). We first examined whether it was possible to decode the abstract patterns (Figure 1A). Decoding analysis was performed using a naïve Bayes implementation of linear discriminant analysis (LDA, Duda et al. 2001), in which the classifier was trained to decode the visual stimulus that a participant was viewing from the corresponding MEG recordings. The decoding analysis was run for all possible pairwise comparisons between visual stimuli for each time point. Figure 2 shows average decoding performance as a function of time. Classification accuracy, reported as *d*-prime, is the average classifier performance. Decoding performance is above chance beginning 40ms after stimulus onset, consistent with estimates of the latency of visual inputs to reach the cortex (Aine, Supek, and George, 1995; Jeffreys and Axford, 1972; Nakamura et al., 1997; Supek et al., 1999), and with the onset of spatial frequency (51ms) and orientation decoding (48-65ms) from MEG (Cichy et al., 2015; Ramkumar et al., 2013). After onset, decoding performance rises to a peak at 90ms and then decays slowly. Following the initial peak at 90ms, there is a second smaller peak in decoding at 400ms, which corresponds to stimulus offset (Carlson, et al. 2011).

Next, we constructed a time-varying RDM from the classification data, which represents the decodability of each stimulus pair as a function of time. Figure 3 shows five frames from the time-varying RDM (see Movie 1 for the complete RDM shown at 5ms resolution). At stimulus onset (0ms), the RDM is dominated by dark blue, indicating a lack of decodability between the stimulus pairs. At 40 ms, which corresponds to the onset of significant decoding performance, a subtle pattern of decodability begins to emerge, reflected in the lighter blue regions of the RDM. At peak decoding (90ms), the RDM is dominated by warm colors, indicating a high level of decodability for most stimulus pairs. The final RDM shown (145 ms) is the time point with the highest correlation with the perceptual RDM (individual subject median: 142.5 ms; group-level: 145 ms).

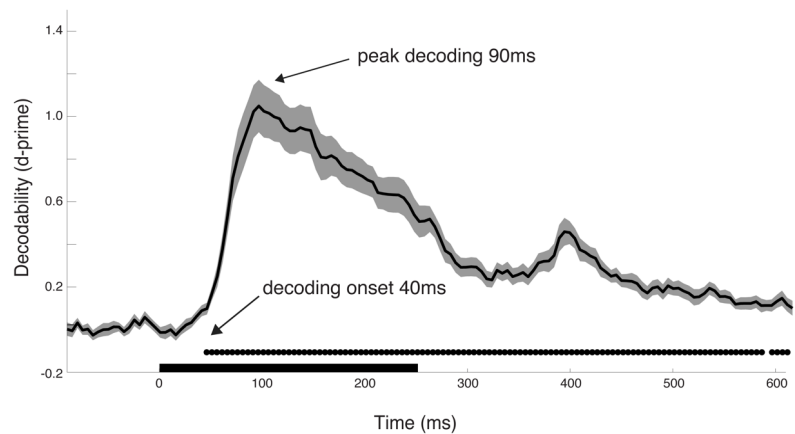

**Figure 2.** Average decodability of all stimulus pairs across time. Solid line is classifier performance (*d*-prime) averaged across all subjects (*N*=20) and stimulus pairs (*N*=325) as a function of time. The black bar on the x-axis corresponds to stimulus presentation (0-250 ms). Shaded region marks +/-1 S.E.M. Disks below the plot indicate above chance decoding performance (onset at 40ms), with significance evaluated using a Wilcoxon signed rank test (FDR < 0.01).

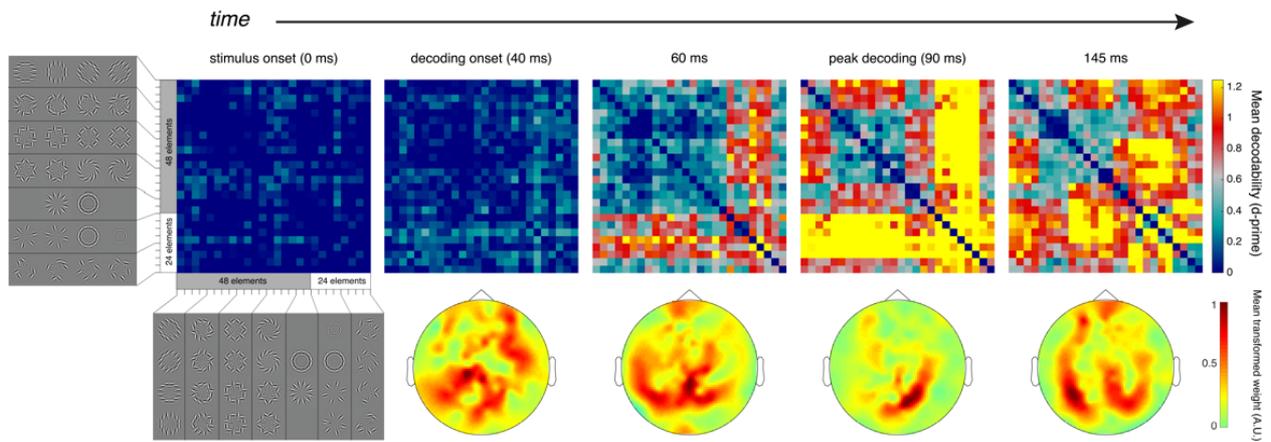

**Figure 3.** Time-varying representational dissimilarity matrix (RDM) for all pairwise stimulus comparisons. Five timepoints are shown, the full time-varying RDM is available online as a movie (see Movie 1). The frames shown here track the evolution in decodability from stimulus onset (0 ms) to peak decoding at 90 ms, which is dominated by warm colors in the RDM, indicating a high level of decodability for most stimulus pairs. Below the RDMs are averaged weight maps in sensor space (averaged across all subjects and stimulus pairs) for four timepoints: 40, 60, 90, and 145ms. Weights are transformed from the classifier output using the method described by Haufe et al. (2014) prior to projection onto sensor space (see Method for further details).

*Perceived Similarity Predicts Decodability*

The capacity to decode the visual stimuli from patterns of neural activation shows that information related to the visual stimuli exists in the MEG signal. We then used RSA to investigate the nature of this decodable signal. The empirical time-varying RDM in Figure 3 (see also: Movie 1) represents the decodability of the neural patterns associated with visual stimulation as a function of time. To summarize the overall decodability of the stimulus set, we calculated the time-averaged RDM from the first time point in which decodability is above chance (40ms) to stimulus offset (250ms). The average RDM (Figure 4A) quantifies how decodable each unique stimulus pair is and measures the similarity between their neural activation patterns. There is clear visible structure in the RDM, indicating that some stimuli share a more similar neural representation than others. The time-averaged RDM in Figure 4A is for illustration, for the formal model comparisons we used the complete time-varying RDM (Figure 3, Movie 1).

Our central question is how perceived similarity relates to the brain's emerging representation of the stimuli. We addressed this within the RSA framework by constructing a perceptual RDM that predicts the relative decodability of each stimulus pair based on perceived similarity as rated by human observers. The perceptual RDM is the average of the normalized ratings for each pair made by each observer (Figure 4B). The perceptual RDM (Figure 4B) shows clear structure, indicating that stimulus pairs varied in their perceived similarity. In order to assess the correspondence between the perceptual RDM and the MEG data, we used a rank-order correlation (Kendall's tau a) between the model and empirical RDMs across time (Figure 4D). Significant correspondence between the model and the data was assessed with a non-parametric Wilcoxon signed rank test.

We observed a strong correspondence between the behavioral ratings of perceived similarity made by human observers and the brain's time-varying representation of the stimuli, which is evident by visual inspection of the neural and behavioral RDMs (compare Figure 4A to the perceptual RDM in Figure 4B). This correspondence is supported by a significant time-varying correlation between the perceptual RDM and decodability (black trace, Figure 4C). The correlation between the model predictions and the decodability of the patterns begins 50ms after stimulus onset, and remains significant over almost the entire time interval. In addition, the correlation between the neural data and perceptual RDM closely tracks the lower bound of the noise ceiling from approximately 150ms after stimulus onset (black dotted line in Figure 4C). This shows that the magnitude of the observed correlation between the behavioral and neural RDMs is within the theoretical upper limits for the data, thus the perceptual RDM provides an explanation of the data comparable with the true (unknown) model (Nili et al, 2014).

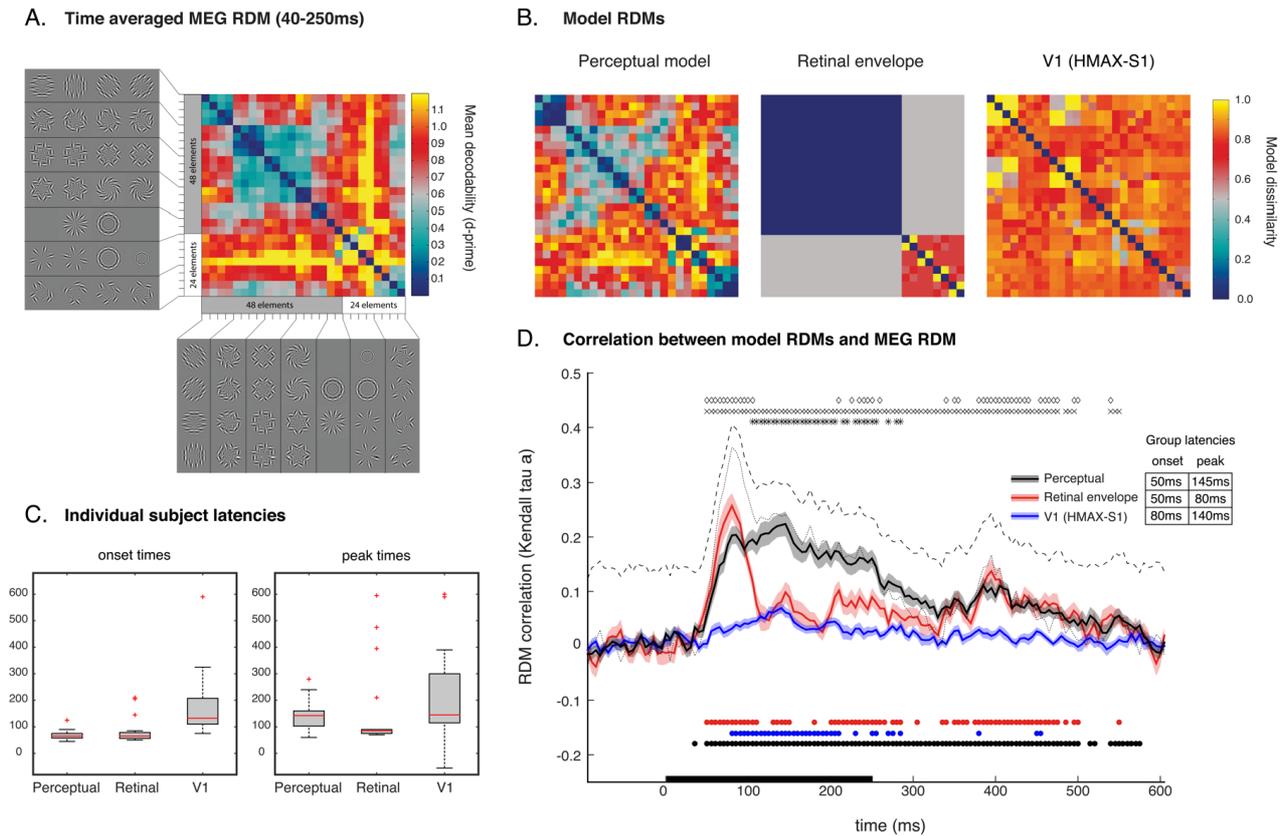

**Figure 4. RDM Model Comparisons.** (A) Empirical RDM displaying the time-averaged decodability of all exemplar pairings from the time decodability first is above chance (40ms) to stimulus offset (250ms). (B) Model RDMs scaled to range from 0 (identical) to 1 (highly dissimilar) for the perceptual similarity model, the retinal envelope model, and the V1 model. Each model makes a prediction for every possible stimulus pairing. (C) Individual subject latencies for the onset of a significant correlation between each of the three models and the MEG RDM (left panel), and the timepoint corresponding to the peak correlation between each model and the MEG RDM (right panel). (D) Group-level correlations between the MEG RDM and each of these three model RDMs. Colored lines are time-varying correlations between model predictions and MEG decoding performance averaged across subjects (shaded region: ± 1 *SEM*). Dashed and dotted lines represent the 'noise ceiling', (Nili, et al., 2014), see Methods for definition and calculation. Colored disks below the plots indicate a significant correlation, evaluated using a Wilcoxon signed rank test (FDR < 0.01). Symbols above the plots indicate a significant difference between the models: diamonds: Retinal envelope vs. V1, crosses (X): Perceptual vs. V1, asterisks (*): Perceptual vs. Retinal envelope. The black bar on the x-axis indicates the stimulus duration (0-250 ms).

*Can Early Visual Representations Explain Decodability?*

Perceptual similarity proved to be a near-optimal model for predicting the neural similarity between abstract visual patterns. For comparison we tested additional models of low-level visual features and early visual processing that we reasoned were likely to predict decodability. First, we constructed a retinal envelope model that predicts decodability based on inter-stimulus differences in retinal projection, as we have previously observed that retinal projection predicts the decodability of higher-level object stimuli from MEG (Carlson et al., 2011). The retinal envelope model (Figure 4B) significantly correlates with the MEG RDM beginning 50ms after stimulus onset (Figure 4D). Following this early onset, the model correlation peaks at 80ms and then declines sharply (these are the group-level latencies, see Figure 4C for the distribution of the latencies for individual subjects). The early success of this model indicates that the difference between the retinotopic projections of stimuli is an important factor in the similarity of their neural representation at the large-scale pattern level, particularly immediately after stimulus onset. The model, however, fails to capture the complex structure of the neural representation of the stimuli (Figure 4A), and following this sharp early peak at 80ms (which is well below the theoretical maximum defined by the noise ceiling), the model's predictive power drops quickly. In addition, the perceptual RDM significantly outperformed the retinal envelope model in explaining the MEG data for a substantial time period following the early peak of the retinal envelope model (Wilcoxon rank sum test with FDR < 0.01, significant time points marked with an asterisk above the plots in Figure 4D).

While retinotopic organization is clearly a dominant organizational principle in visual cortex, early visual areas also encode a range of visual features, e.g. orientation, that are not present in the retinal envelope model. Orientation selectivity is evident in the earliest stages of visual processing and is encoded in simple cell neurons in visual cortex (Hubel and Wiesel, 1962, 1968). To construct a more complete model of early visual processing, we built a model based on the response properties of V1 simple cells from the predicted response profiles to the visual stimuli from the output of the S1 layer of HMAX, a computational model of early visual processing, which represents orientation at multiple scales (Riesenhuber and Poggio, 1999; Serre and Riesenhuber 2004; Hubel and Wiesel, 1965). The V1-HMAX model in Figure 4B predicts that nearly all stimulus pairs will be highly decodable. The model did fit the MEG data beginning from 80ms, with a peak at 140ms. However, the V1-HMAX model did not approximate the noise ceiling at any latency, and was not as strong a predictor of the neural data as either the perceptual RDM or the retinal envelope model. The difference between models was significant, both the retinal envelope model and the perceptual RDM had a significantly larger correlation with the MEG RDM than the V1 model (significant time points are marked by diamonds and crosses respectively above the plots in Figure 4D). Additional "higher" layers of HMAX up to layer C2 were also tested and performed similarly (Figure 5). Each layer of HMAX first reached a significant correlation with the empirical MEG RDM between 55-90ms, and for some sporadic time points thereafter, but overall none of the HMAX layers was a strong predictor of the MEG data.

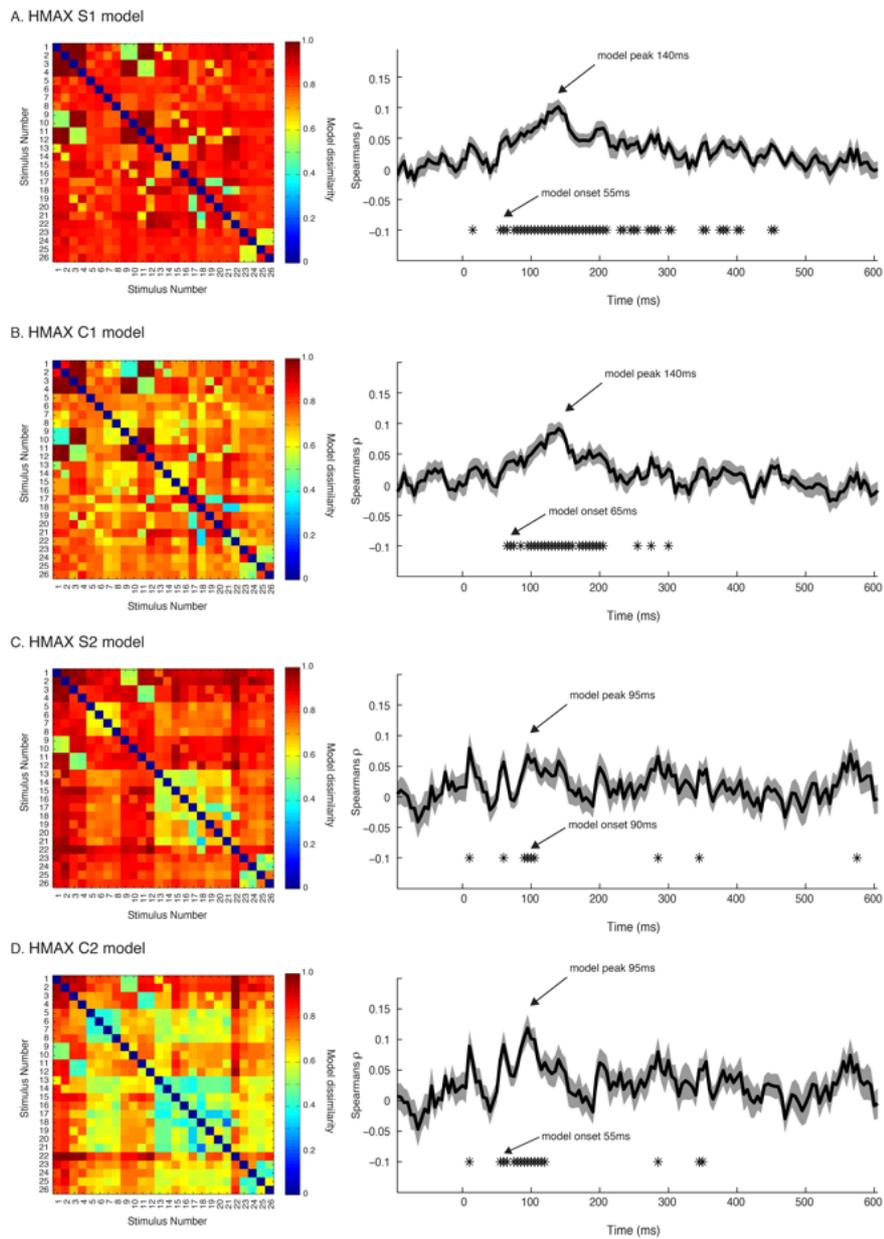

**Figure 5. Average decodability across four layers of HMAX.** Each row A-D shows the model predictions (left panel) and the time-varying correlation between the model and the MEG data (right panel) separately for a level of HMAX (S1, C1, S2, C2). Panels on the left show the model RDM's for the four HMAX layers. Color values in the RDM represent the dissimilarity between the pairs of patterns as predicted by the assumptions of each model layer. Panels on the right show the model correlation with MEG decoding performance. Plotted is the Spearman rank-order correlation between the model RDM and the time-varying MEG decoding RDMs. The solid line is the average correlation across subjects. The shaded region is +/- 1 S.E.M. Asterisks below the plot indicate a significant correlation, evaluated using a Wilcoxon signed rank test (FDR < 0.01).

We speculated that one reason for the limited explanatory power of the V1 model based on HMAX is possibly because it assigns too high a weight to local orientation differences between the stimuli, and fails to capture the perceptually salient differences in global form, which are highly weighted by the perceptual RDM. To verify that local orientation differences are a poor predictor of decodability, we constructed a RDM based on the overall magnitude of the orientation disparity between corresponding elements in the stimulus pairs (Figure 6A). Although this model was unsuccessful at predicting the neural data at any time point (Figure 6A), we found that we could decode the orientation of the stimulus pairs that had a coherent global orientation (Figure 6B, blue trace; Figure 1, Set 1), consistent with previous reports of orientation decoding with MEG (Duncan et al, 2010; Ramkumar, et al. 2013). Analogous to fMRI results (Alink et al., 2013), decoding was moderated by orientation coherence among the elements, because stimulus pairs with an equivalent local disparity but an absence of a coherent global orientation could not be decoded (Figure 6B, green trace; Figure 1, Set 2). Although only coherent stimuli could be decoded, the difference between the coherent and incoherent waveforms only reached significance at sporadic time points (Wilcoxon rank-sum FDR $p<0.01$, asterisks above the plot in Figure 6B), suggesting that incoherent stimuli could probably be decoded with increased statistical power (for example, in a comparable experiment with less exemplars). However, when considered in conjunction with the comparisons made below for stimuli differing in global shape, this pattern of results suggests that grouping across local elements may be an important component of the underlying neural representation.

Orientation decoding with fMRI has been suggested to be a byproduct of the radial bias – the greater number of neurons representing orientations pointing toward the fovea (Levick and Thibos, 1982; Leventhal and Schall, 1983; Schall et al., 1986; Sasaki et al., 2006; Mannion et al., 2010), however, this issue remains controversial (e.g. Carlson, 2014; Freeman et al., 2011; Freeman et al., 2013; Mannion et al., 2009; Alink et al., 2013; Maloney, 2015; Clifford and Mannion, 2015; Carlson and Wardle, 2015). We found that a RDM modeled on inter-stimulus differences in the radial bias did not fit the MEG data; the radial preference model never reached significance at any time point (Figure 6A). In addition to the failure of the radial preference model, we found that decoding of stimulus pairs designed to test for radial bias effects was instead moderated by differences in their global form. Stimuli that were matched for the magnitude of the radial bias that had similar global form (within-shape decoding of stars or spirals; see radially balanced pairs in Figure 1, Set 4) could not be decoded (Figure 6C, green trace). However, these radially-balanced stimuli could be decoded in between-shape pairs (i.e.: across shape decoding of stars versus spirals) in which they differed in global form (Figure 6C, blue trace). The difference between the within-and between-shape conditions was significant for the majority of the stimulus duration (Wilcoxon rank-sum FDR $p<0.01$; significant timepoints marked by asterisks (*) above the plot in Figure 6C). Furthermore, the 'opposite bias' stimulus pair that was maximally different with respect to the radial bias (strong [spokes] versus weak [rings], see final pair in Set 4, Figure 1) could be decoded (Figure 6C, black trace), and was significantly different than the within-shape pair for most of the stimulus duration (diamonds above plots in Figure 6C). However, decoding performance for the opposite bias pair was not substantially better than the between-shape pair that were radially balanced but differed in global form (crosses above Figure 6C, only sporadic time points are different). As only the between-shape and opposite-bias pairs (that also differed substantially in global form) were decodable, these results may be interpreted as additional support for the importance of global form in the neural representation.

Although differences in the radial bias did not appear to modulate decodability in our stimulus set, it is established that radially balanced patterns can be decoded from fMRI (e.g. Mannion et al, 2009; Freeman et al., 2013; Alink et al, 2013). We speculate that the reason we were unable to decode radially balanced spirals is likely because we are decoding whole-brain MEG activation patterns for a relatively large stimulus set ($n$=26). Decoding of spirals with fMRI has been done from isolated activity in visual cortex and a small number of stimuli ($n$=2-8). Consistent with this explanation, radially balanced spirals have recently been decoded with MEG as part of a smaller stimulus set ($n$=4 stimuli) (Cichy et al., 2015).

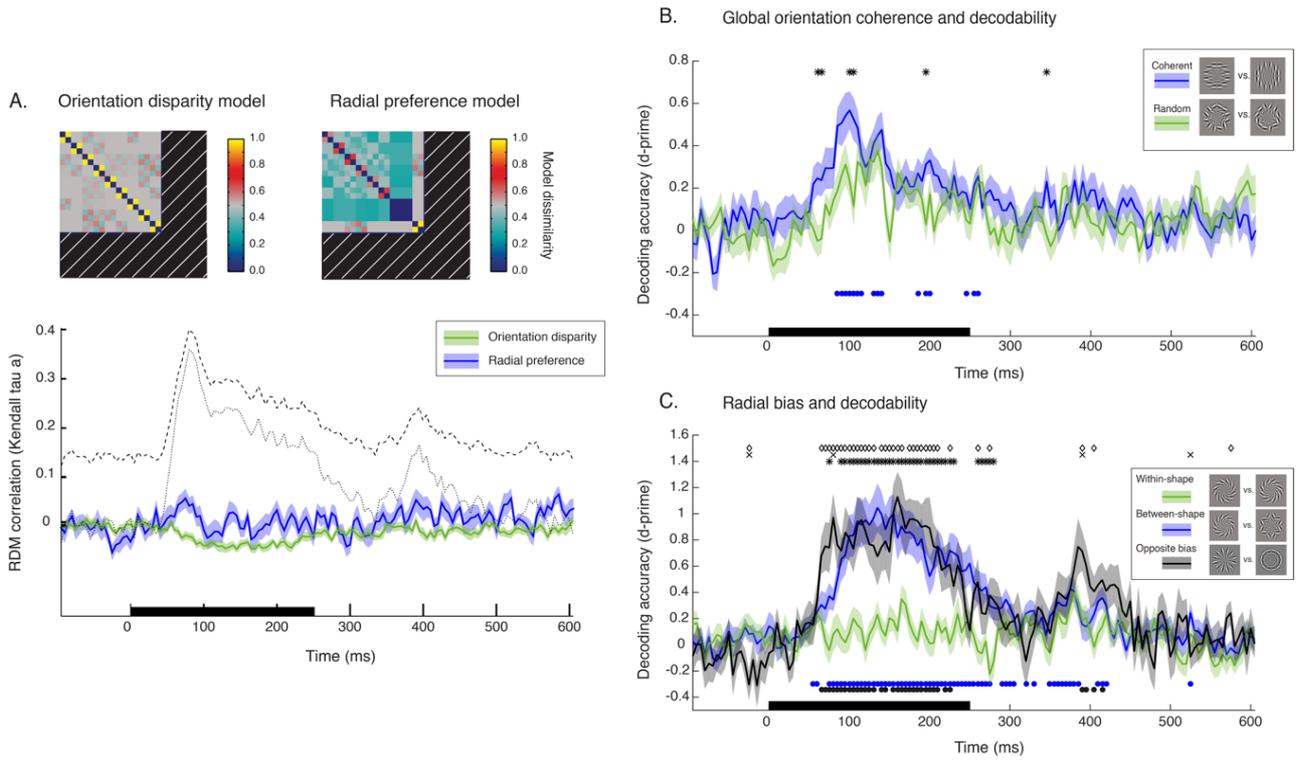

**Figure 6.** Orientation and the radial bias. (A) Top: Model RDMs for orientation disparity and radial preference. Hatched regions mark the undefined predictions for each model; both orientation disparity and radial preference were only calculated for the patterns with all 48 elements in corresponding retinal locations. Bottom: Correlation between model RSMs and the MEG RDM (details as in Figure 4D). (B) Orientation: Average decodability for all pairwise comparisons (*n*=6) between the four patterns that have a coherent global orientation (blue trace), and average decodability (*n*=6) for the four 'random' patterns which have equivalent local orientation disparity without coherent global orientation (green trace). (C) Radial bias: Average decoding accuracy for the two radially balanced pairs of the same shape - stars or spirals (green trace); average decoding accuracy for the four possible between-shape pairs (blue trace); and decodability for the stimuli differing in the strength of the radial bias (black trace). Errors are ± 1 *SEM*. Colored discs below each plot indicate time points with significant decoding (matched to color of individual traces). Symbols above each plot indicate a significant difference between conditions at that timepoint: diamonds: Within vs. Opposite, crosses (X): Between vs. Opposite, asterisks (*): Between vs. Within.

**Discussion**

Our main finding is that the perceived similarity of visual patterns predicts their representational similarity in whole-brain neural activation patterns measured with MEG. We observed that perceptual similarity ratings reached the limits of the highest possible correlation with the representational structure measured with MEG as early as 150ms after stimulus onset, and the success of the model persisted for several hundred milliseconds beyond stimulus offset. This demonstrates that differences in perceived global form are matched by equivalent differences in neural representational distance. The perceptual RDM based on human ratings of similarity reached the lower bounds of the 'noise ceiling' (Nili et al, 2014), which indicates that the perceptual RDM explained as much of the variability in the similarity of the brain activation patterns elicited by the visual stimuli as the unknown true model. The noise ceiling provides a guide for settling on a satisfactory model within the RSA framework: when the bounds of the noise ceiling are reached it indicates that the model provides as complete an explanation for the data as is possible within the limits set by the noise in the data.

Previously, two computational models have been reported which reach or closely approximate the noise ceiling. A computational model based on a supervised deep convolutional neural network reached the lower bounds of the noise ceiling for explaining fMRI activation patterns for a diverse set of objects in human IT (Khaligh-Razavi & Kriegeskorte, 2014). Similarly, a biologically-plausible hierarchical convolutional neural network model approached the lower limits of the noise ceiling for neural data from monkey IT in response to a large set of object stimuli (Yamins et al., 2014). Building on the success of these computational models, the perceptual RDM reported here is the first behavioral model to our knowledge within the RSA framework that reaches the noise ceiling. The correlation we observe between behavioral similarity ratings and MEG activation patterns is consistent with several earlier studies that have observed a correspondence between behavioral ratings and fMRI activation patterns (e.g. Edelman et al 1998; Mur et al 2013; Connolly et al. 2012; Hiramatsu et al 2011; Op de Beeck et al., 2008).

The strong correspondence we observed between behavior and neural representation is a reflection of our stimulus set, which was designed to probe the neural representation of global form while controlling for low-level visual features. As all stimuli were constructed from identical visual features (Gabor patches), we assume that observers based their similarity judgments on the overall global form or Gestalt of each pattern created by the particular arrangement of Gabor patches. The fact that global form is the most salient difference between our stimuli is also consistent with the relatively poor performance of the V1 model based on HMAX. We suggest that the poor performance of the V1 model is likely because it weights local orientation differences highly while ignoring global form, and local orientation differences were a poor predictor of decodability. The best performing model assessed using RSA is always relative to the stimulus set, thus in order to demonstrate a tight link between perceptual similarity and neural activation patterns, it is necessary to use stimuli in which differences in global form are separated from both semantic similarity and low level visual parameters.

Similarly, Mur et al. (2013) used RSA and found that human similarity judgments for higher-level object stimuli did show similarity with categorical divisions in representational structure in IT; however, in this case the similarity judgments contained additional structure not present in the neural representation. The human judgments showed a tighter categorical clustering than the fMRI data, and contained a finer grain of categorical distinctions. In the Mur et al. (2013) study, similarity judgments were likely based on both semantic and visual characteristics, as the stimuli were pictures of objects, which have inherent conceptual meaning. The visual stimuli we used were abstract, thus we assume observers' similarity judgments were based solely on perceived visual similarity.

Overall the perceptual RDM provided the best explanation of the variance in the MEG data, however; early after stimulus onset (50ms), a simple model based on retinal stimulation predicted decodability as well as the perceptual RDM for a short time window (approx. 50 ms) before its performance fell. We interpret the steep simultaneous rise in explanatory power of the retinal envelope model and the perceptual RDM soon after stimulus onset as a reflection of the overlap between low-level stimulus similarity and perceptual similarity. It is intuitive that low-level features of abstract visual patterns such as retinotopic stimulation, and local regions of luminance and contrast, are part of what makes stimuli appear perceptually similar to human observers The ratings of perceptual similarity made by human observers can be thought of as a 'shortcut' to identifying the perceptually relevant stimulus features that are important in the neural representation. Our data is consistent with previous fMRI results, which show that the relationship between perceived similarity and the similarity of activation patterns cannot be completely explained by similarities in retinal stimulation. Op de Beeck et al. (2008) controlled for retinal envelope by constructing novel objects that varied systematically in both their overall shape envelope (e.g. tall vs. long) and their shape (e.g. sharp vs. curved edges). Notably, Op de Beeck et al. (2008) jittered the retinal position of their shape stimuli, which was constant in our study. Op de Beeck et al. found that fMRI activation patterns for novel objects in LOC were more correlated with perceived shape similarity (e.g. sharp vs. curved edges) than with the similarity in their shape envelope. This is also consistent with fMRI adaptation to stimuli of the same shape that have different local contours (Kourtzi & Kanwisher, 2001).

Our finding that whole-brain activation patterns reflect perceptually important features is consistent with recent neurophysiological and neuroimaging studies suggesting that the representation of visual inputs changes throughout the visual stream. These studies have shown that the representation in early visual areas reflects low-level visual features such as image statistics (Clarke and Tyler, 2014; Hiramatsu, et al 2011). In higher visual brain regions, the representation is instead based on higher-level features such as object category membership (Edelman et al., 1998; Clarke and Tyler, 2014), perceived face identity (Rotshtein et al, 2005), or shape similarity (Kourtzi & Kanwisher, 2001; Op de Beeck et al., 2001, 2008; Haushofer et al., 2008). Furthermore, differences in image statistics are diagnostic of the degree of dissimilarity of large-scale activation patterns measured with EEG (Groen et al., 2012). The early success of the retinal envelope model in predicting the decodability of our stimuli (peak performance just 80ms after stimulus onset) is consistent with the dominance of early visual features (such as contrast) in the representational structure directly after stimulus onset, which later evolves into a representation highly correlated with perceptual similarity.

Although the perceptual RDM was a strong predictor of the MEG data, the behavioral similarity ratings and the MEG data were collected from independent groups of subjects, and the behavioral task involved a relatively coarse judgment of similarity for each pair of stimuli on a scale from 1 to 100. The use of separate subjects for neural and behavioral data collection is common in RSA studies, and a strength of the RSA approach is its ability to examine representational structure across different subjects and methodologies. Future work will determine whether this can be achieved at finer scales (for example, fine perceptual discriminations). A further implication of the success of the behavioral data in predicting the neural representation (from a separate pool of subjects) is that it provides empirical validation of the common assumption that the structure of brain representations can be inferred from behavioral research. Individually the behavioral and MEG study would have reached the same conclusion; however, the bridging of the two studies using the RSA framework strengthens the conclusion and validates each approach (behavior and neuroimaging).

A recent paper in *NeuroImage* also decoded gratings from MEG activation patterns (Cichy et al., 2015), however; the authors draw different conclusions. In fMRI, a decade-long debate over the information source underlying orientation decoding from patterns of BOLD activation persists (e.g. Kamitani & Tong, 2005; Haynes & Rees, 2005; Mannion et al., 2009; Swisher et al, 2010; Freeman et al., 2011; Freeman et al., 2013; Alink et al., 2013; Carlson, 2014; Maloney, 2015; Clifford and Mannion, 2015; Carlson and Wardle, 2015). As orientation columns are at a finer scale than fMRI voxels, it has been suggested that orientation decoding from V1 is evidence that finer-scale information can be accessed at the level of cortical columns with multivariate fMRI (Kamitani & Tong, 2005). Alternatively, others have argued that orientation decoding with fMRI can be explained by coarse-scale biases across voxels (Freeman et al., 2011, 2013) or edge-related activity (Carlson, 2014). Cichy et al. (2015) decoded gratings from MEG activation patterns using several stimulus controls, and in conjunction with theoretical simulations to demonstrate technical plausibility, conclude that it is likely that the decodable orientation information in their MEG signals originates from the spatial scale of cortical columns. In contrast to BOLD activation, MEG activation patterns cannot be unambiguously spatially localized to early visual cortex, adding a level of difficulty to identifying the spatial scale of the decodable MEG signal. Cichy et al. (2015) suggest that the early onset of orientation decoding (~ 50ms) in their experiments is consistent with a locus in early visual cortex; however, we also observe an early onset of decoding (~ 40ms) and found that the perceptual RDM significantly correlated with the MEG RDM as early as 50ms. As our results underscore the importance of global form and perceptual similarity in the decodable MEG signal (in contrast to the relative underperformance of the orientation and V1 models), we suggest caution in interpreting the source of decodable orientation information in MEG signals as originating from the scale of cortical columns.

**Conclusions**

We found that visual stimuli that were perceived to look more similar to each other by human observers also had more similar complex neural activation patterns as measured with MEG. The behavioral model was a near-optimal predictor of neural representational similarity, and closely tracked the maximum possible correlation with the neural data from just 150 ms post-stimulus onset. The results show that the perceptual Gestalt of an image is captured in coarse-scale neuromagnetic activation patterns, and thus provide evidence that perceived similarity can indeed be conceptualized as representational distance. The decodable MEG signal emerges from complex neural activity at multiple scales throughout the visual processing hierarchy, and it is both remarkable and logical that the representational geometry of this pooled neural activity represents an end-stage as advanced as human judgments of perceived similarity.